\documentclass[proof]{pasj00}
\Received{$\langle$ $-$ $\rangle$}
\Accepted{$\langle$ $-$ $\rangle$}
\Published{$\langle$publication date$\rangle$}
\SetRunningHead{Suzuki,~T. et al.}{Spatial distribution of cold and warm
dust in M101} 

\usepackage{times}

\begin{document}

\title{Spatial Distributions of Cold and Warm Interstellar Dust in M101
\\Resolved with AKARI/Far-Infrared Surveyor (FIS)} 
\author{Toyoaki \textsc{Suzuki}, Hidehiro \textsc{Kaneda}, Takao
\textsc{Nakagawa}, Sin'itirou \textsc{Makiuti}, Yoko \textsc{Okada}}
\affil{%
   Institute of Space and Astronautical Science, Japan Aerospace
   Exploration Agency, 3--1--1 Yoshinodai,\\
   Sagamihara-shi, Kanagaawa 229--8510}
\email{suzuki@ir.isas.jaxa.jp}
\author{Hiroshi \textsc{Shibai}, Mitsunobu \textsc{Kawada}}
\affil{%
   Graduate School of Science, Nagoya University, Furu-cho,\\
   Chikusa-ku, Nagoya 464--8602}
\and
\author{Yasuo \textsc{Doi}}
\affil{%
   Department of Earth Science and Astronomy, University of Tokyo, 3--8--1 Komaba,\\
   Megro-ku, Tokyo 153--8902}


\maketitle

\begin{abstract}
 The nearby face-on spiral galaxy M101 has been observed with the
 Far-Infrared Surveyor (FIS) onboard AKARI. The far-infrared four-band
 images reveal fine spatial structures of M101, which include global
 spiral patterns, giant H\emissiontype{II} regions embedded in outer
 spiral arms, and a bar-like feature crossing the 
 center. The spectral energy distribution of the whole galaxy shows the
 presence of the cold dust component (18$^{+4}_{-10}$ K) in addition to
 the warm dust component (55$^{+9}_{-25}$ K).  
 The distribution of the cold dust is mostly concentrated
 near the center, and exhibits smoothly distributed over the entire
 extent of the galaxy, whereas the distribution of the warm dust
 indicates some correlation with the spiral arms, and has spotty
 structures such as four distinctive bright spots in the outer disk in
 addition to a bar-like feature near the center tracing the CO intensity
 map. The star-formation activity of the giant H\emissiontype{II}
 regions that spatially correspond to the former bright spots is found
 to be significantly higher than that of the rest of the galaxy. The
 latter warm dust distribution implies that there are significant
 star-formation activities in the entire bar filled with molecular
 clouds. Unlike our Galaxy, M101 is a peculiar normal galaxy with
 extraordinary active star-forming regions.    
\end{abstract}

\section{Introduction}
The characteristics of large-scale star formation are of great importance
to understand the evolution of a galaxy. One of the open fundamental 
questions is the dependence of the global star formation rate (SFR) 
on the total gas content of a galaxy. Far-infrared (IR) dust emission 
can provide us with reliable estimates of SFR. 
In general, spiral galaxies have cold and warm dust, as first suggested
by \citet{jong}, and confirmed by ISO observations (see review of
\cite{sauvage} and references therein). The cold dust component
($T_{\rm{d}}$ = 10--20 K) is associated with molecular 
and atomic hydrogen clouds, which is heated mostly by the general
interstellar radiation field (ISRF) \citep{cox}. The cold dust 
accounts for more than 90 \% of the total interstellar dust in mass, and
can therefore be used to estimate the total gas distribution over a
galaxy. The contribution of the warm dust component ($T_{\rm{d}}$ =
25--40 K) can be used to estimate the SFR of massive stars ($\gtrsim$4
\MO); since the warm dust component is heated by O and early B stars in
H\emissiontype{II} regions \citep{cox}, its luminosity can reasonably be
considered to reflect the SFR of massive stars, providing that the
escape fraction of non-ionising UV photons from the HII regions is
properly taken into account (\cite{pope2, hippelein3}). Hence, to
properly address the question about the relation between the global SFR
and the total gas content in spiral galaxies, it is crucial to separate
the contributions of the cold and warm dust components and to discuss
the spatially-resolved distribution of each component within a
galaxy. Particularly for late-type spiral and irregular galaxies, most
of the far-IR luminosity is carried by the cold dust primarily emitting
longwards of the IRAS limit of 120 $\mu$m, and thus observations at
wavelengths longer than 120 $\mu$m are essential to detect the cold dust
component (\cite{pope1, val}).  

Here we present new far-IR images of the nearby galaxy M101 obtained with the 
Far-Infrared Surveyor (FIS; Kawada et al. 2007) onboard AKARI (Murakami et al. 2007).
M101 is a face-on spiral galaxy with global spiral patterns, 
classified as Sc(s)I \citep{sandage} with a distance of 7.4 Mpc \citep{jurcevic}.  
The galaxy is an excellent candidate for this study, since
it has a large optical size of 28$\times$28 arcmin$^2$
\citep{Nilson}, well-developed spiral arms, and several conspicuous giant
H\emissiontype{II} regions. \citet{rice} and \citet{nicholas} presented
the 60 and 100 $\mu$m images of M101 with IRAS; the latter found spatial
correspondence between the morphology of far-IR and H$\alpha$
luminosities. However, since the angular resolution of IRAS is rather poor,
structures such as spiral arms and star-forming regions cannot be
distinguished. As an early release of ISO observations, Hippelein et
al.~(1996a, 1996b) presented the 60, 100, and 170 $\mu$m images of M101
with ISOPHOT, which resolved the giant H\emissiontype{II} regions. The
better 100 $\mu$m image was later derived by \citet{tuff}. The far-IR
colors are found to be surprisingly insensitive to the intensity of the
ISRF except for the two giant H\emissiontype{II} regions, NGC 5447 and
5461. The SEDs for the total galaxy and also for local specific fields
show that a single blackbody model with temperature of about 30 K can
fit the observed SEDs.  

\citet{siebe} observed 16 galaxies including active (Seyfert and
starburst) and inactive (normal) galaxies. The SEDs of the active
galaxies can be described by a single blackbody model at temperatures of
31.5$\pm$2.8 K. The ratio of the far-IR luminosity to the gas mass,
$L_\mathrm{FIR}/M_\mathrm{gas}$ is $\sim$90 \LO/\MO. In contrast, the
SEDs of the inactive galaxies require the presence of cold dust at
temperatures of 12.9$\pm$1.7 K in addition to warm dust at 31.8$\pm$2.8
K, while $L_\mathrm{FIR}/M_\mathrm{gas}$ is $\sim$3 
\LO/\MO. From the IRAS and ISO observations, M101 is classified as an active
galaxy in terms of its SED, however it is an inactive galaxy considering its 
small $L_\mathrm{FIR}/M_\mathrm{gas}$ value (0.5 \LO/\MO) with the far-IR 
luminosity of 1.1$\times$10$^{10}$ \LO \citep{rice} and the gas mass
of 2.4$\times$10$^{10}$\MO \citep{kenney, allen}. Hence the activity of M101
 is still controversial.

A remarkable property of M101 is its large, bright, and metal-poor 
H\emissiontype{II} regions (NGC 5447, 5455, 5461, 5462, and 5471) located 
in the outer disk of the galaxy. To explain the properties of the giant
H\emissiontype{II} regions, \citet{kenney} proposed that either the
initial mass function is unusually enhanced in massive stars or the gas is
consumed up efficiently in these regions. For the former, \citet{rosa}
concluded that the initial mass function for the stars with the masses larger
than 2 \MO is rather normal in M101, similar to that in the solar 
neighborhood. For the latter, there was an indication that the star 
formation efficiency (SFE) of the massive stars in NGC 5461 is higher 
than a typical SFE observed in our Galaxy \citep{blitz}. 
\citet{jean} detected CO emission from two giant
H\emissiontype{II} regions (NGC 5461 and 5462); for NGC 5461, they found
that the SFE is unusually high as compared with that in star-forming
regions of our Galaxy, concluding that higher SFE is likely to be the key to
the formation of giant H\emissiontype{II} regions in M101. Comparative
research would however be necessary over the entire galaxy, not just
restricted to the giant H\emissiontype{II} regions.  

The far-IR four bands of the FIS have a great advantage over IRAS and 
Spitzer/MIPS in detecting both cold and warm dust components, which offers
 a unique capability to spatially and spectrally separate the two dust 
components determining the distribution of star-formation activity over
a galaxy. The high spatial resolution of the FIS has an advantage over
IRAS and ISO in resolving spiral arm and inter-arm regions, and
identifying H\emissiontype{II} regions embedded in the
arms. Furthermore, a large dynamic range in the signal detection is
another advantage over MIPS; the FIS can observe a galaxy  
without being prevented by saturation effects at the brighter center and 
the H\emissiontype{II} regions. Therefore, AKARI/FIS observations are best 
suited for the study of the luminous galaxy M101.  

\section{Observations and results}
The observations were performed as part of the FIS calibration program  
on June 14th in 2006 by using one of the FIS observation modes, FIS01. 
The FIS was operated in a photometry mode with the four bands: N60
(65 $\mu$m), WIDE-S (90 $\mu$m), WIDE-L (140 $\mu$m), and N160 (160
$\mu$m). The observations consist of two sets of round-trip slow scans with a
shift in the cross-scan direction. The round-trip scan ensures data
redundancy for the corrections of radiation effects. The user-defined parameters are 
the scan speed of 8 arcsec/sec, the cross-scan shift length of 240 arcsec, 
and the reset time interval of 1.0 sec. 

The FIS01 scan sequence is shown in figure 1, which consists of (1) the
first round-trip, (2) the cross-scan step, and (3) the second round-trip. In
figure 1, the area inside the thick lines is scanned in the four bands.  
The total area scanned in the four bands is 15$\times$12 arcmin$^2$.
Details of the FIS instrument and its in-orbit performance/calibration
are described in Kawada et al. (2007).

The FIS data were basically processed with the AKARI official pipe
line modules. In addition to these, we applied a series of corrections
for the radiation effects developed in \citet{suzuki}. The
long-term (several times $10^3$ seconds) gradual changes in the detector 
responsivity due to passage of the South Atlantic Anomaly were 
corrected by using a set of the internal calibration signals. 
Then, cosmic-ray glitches were detected and their effects were corrected;
spikes caused by ionizing radiation hits were removed and base-line 
fluctuations due to short-term (a few seconds) changes in the detector
responsivity preceded by the spikes were restored at each detection
point. The distortion of the fields-of-view of the FIS array detectors 
(SW and LW; Kawada et al. 2007) and their alignment was corrected. 
The positional offset uncertainty between the FIS detectors was
estimated to be about 10 arcsec from the observations of point sources  
such as far-IR-bright asteroids. The focal-plane coordinate system was 
then converted into the equatorial coordinate system. Finally, the four-band 
images were created with grid sizes of 25 arcsec for the WIDE-L and N160
bands and 15 arcsec for the WIDE-S and N60 bands. The widths (FWHM) of
the Point Spread Functions (PSFs) are $\sim$60 arcsec for the WIDE-L and
N160 bands and $\sim$40 arcsec for the WIDE-S and N60 bands (Kawada et
al. 2007).  

The far-IR flux densities in the four bands were obtained by integrating the 
surface brightness within an aperture after subtracting the sky background
level. The background levels were estimated from nearby blank skies and
subtracted from images, which were observed in the beginning and at the
end of the scan observation. To obtain the fluxes of the whole galaxy,
the photometric aperture with a radius of 8 arcmin around the center was
used, which was large enough to cover a dominant fraction of the far-IR
emission from the galaxy without degrading the S/N. On the basis of the
brightness profile along the scan direction, the loss of fluxes outside
the aperture was estimated to be less than 20 \%. To obtain the fluxes
in image bins (14 arcsec radius; see below) that are significantly
smaller than the FIS beam sizes, appropriate aperture corrections were
applied to the four-band data, as described in \S3.2.  
Color corrections were applied for the obtained flux densities.
The resultant correction errors are estimated to be $\sim$30 \% for the
WIDE-L band, $\sim$40 \% for the N160 band, and $\sim$20 \% for the WIDE-S 
and N60 bands under the current stage of the FIS calibration,
which are expected to improve with progress in the calibration.      
Table~\ref{flux} shows the derived flux densities of M101 in the four
bands of the FIS. 

Figure~\ref{4image} represents the far-IR four-band images of M101.
The images are smoothed with boxcar kernels with a width of 50 arcsec 
for the WIDE-L, N160 bands and 30 arcsec for the WIDE-S, N60 bands.   
At wavelengths longer than 120 $\mu$m, the WIDE-L and
N160 band images clearly exhibit bright spots embedded in the arms in
addition to the spiral patterns as compared to the ISO 170 $\mu$m
image (\cite{hippelein}). The conspicuous four bright spots spatially correspond 
to the four giant H\emissiontype{II} regions, NGC 5447, 5455, 5461, and 5462, 
and they are even more emphatic in the WIDE-S and N60 band images. Among them, 
NGC 5461, which is located at 4.5 arcmin south-east from the center is 
brightest and even brighter than the center. Furthermore, as seen in the
ISO 100 $\mu$m image (\cite{tuff}), a bar-like feature crossing the center can also
be seen in the WIDE-S and N60 band images, which have striking
resemblance to the distribution of CO emission \citep{kenney}.

\section{Discussion}

\subsection{Presence of cold dust component}
The spectral energy distribution (SED) of the whole galaxy
is obtained by integrating the surface brightness as described in \S2.
Figure~\ref{total_sed} shows the resultant SED of M101. Integrated flux
densities in the four bands are shown by filled  
boxes, while those in the far-IR bands of ISOPHOT at 60, 100 and 170 $\mu$m
obtained with the latest calibration (\cite{tuff}) are shown by
open boxes. As seen in the figure, our results with the
AKARI/FIS are consistent with the newly-calibrated ISOPHOT data.  
It is clear from the figure that a single-temperature blackbody spectrum
cannot reproduce the observed SED. We have therefore fitted the 
AKARI and ISO data with a double-temperature blackbody model modified by
an emissivity power-law index of 1:     
\begin{equation}
 F_\mathrm{FIR}(\nu) =
 A_\mathrm{C} \nu \pi B_{\nu}(T_\mathrm{c}) + A_\mathrm{W} \nu \pi B_{\nu}(T_\mathrm{w}),
\end{equation}
where $T_\mathrm{c}$, $T_\mathrm{w}$, $B_{\lambda}(T)$, $A_\mathrm{C}$,
and $A_\mathrm{W}$ are the temperatures of the cold and warm dust, the
Planck function, the amplitudes of the cold and warm dust components, respectively.  
The best-fit model thus obtained is indicated by
the solid line in figure~\ref{total_sed}, while the dotted line and the dash-dotted
line represent the warm and the cold dust component, respectively. The 
best-fit temperatures representative of $T_\mathrm{c}$ and 
$T_\mathrm{w}$ are 18$^{+4}_{-9}$ K and 55$^{+9}_{-25}$ K, 
respectively. The errors of the dust temperatures are derived from
the 1$\sigma$ confidence contour ( $\Delta \chi^2$ = 2.3 ) encompassed by 
the two parameters ($T_\mathrm{c}$, $T_\mathrm{w}$), while the other 
parameters ($A_\mathrm{C}$, $A_\mathrm{W}$) are fixed at the best-fit values. 
For the error in the warm dust temperature,
since the upper error limit is not well determined by the data at wavelengths
longer than 60 $\mu$m alone, the error is limited by combining the IRAS flux density at
25 $\mu$m \citep{rice}. If we apply an emissivity power-law index of 2, 
the cold and warm temperatures to explain the resultant SED are 
15$^{+2}_{-7}$ K and 37$^{+16}_{-11}$ K, respectively, which do not make 
significant differences from the above temperatures.
\citet{siebe} showed that inactive
spiral galaxies possess cold dust of temperatures of 12.9$\pm$1.7 K and
warm dust of 31.8$\pm$2.8 K on the average, which are compatible with
the two dust temperatures obtained for M101 with the FIS. Hence, we have 
clearly confirmed the presence of the cold dust component in M101.  

By using the best-fit double temperature modified blackbody model, the far-IR
luminosity of the cold dust component, $L_\mathrm{C}$, and the warm dust
component, $L_\mathrm{W}$, can be calculated as follows: 
\vspace{1mm}
\begin{equation}
 L_\mathrm{C} = 4\pi D^2 A_\mathrm{C} \int \nu \pi B_{\nu}(T_\mathrm{c}) \mathrm{d}\nu
\end{equation}
\vspace{-3mm}
\begin{equation}
 L_\mathrm{W} = 4\pi D^2 A_\mathrm{W} \int  \nu \pi B_{\nu}(T_\mathrm{w}) \mathrm{d}\nu,
\end{equation}
\vspace{1mm}
where $D$ is the distance to M101 (7.4 Mpc; \cite{jurcevic}). From
table~\ref{dmasL}, the resultant total far-IR 
luminosity $L_\mathrm{FIR}$ (=$L_\mathrm{C}+L_\mathrm{W}$) is (2.1$^{+0.5}_{-0.4}$)$\times$10$^{10}$
$L_\odot$. The errors of the luminosities come from those of the
temperatures and the amplitudes of the two dust components. Hence,
$L_\mathrm{FIR}/M_\mathrm{gas}$ is estimated to be 0.9 $L_\odot/M_\odot$
with the total gas mass ($M_\mathrm{H_2}$+$M_\mathrm{H_I}$) of 2.4$\times$10$^{10}$ 
$M_\odot$ \citep{kenney, allen}. By taking into account the presence of the cold
dust component as well as the small $L_\mathrm{FIR}/M_\mathrm{gas}$
value, M101 can be classified as an inactive galaxy. 

The masses of the cold and warm dust components are estimated to derive
the gas-to-dust ratio over the entire galaxy. 
We applied the grain emissivity factor given by \citet{hildebrand},
the average grain radius of 0.1 $\mu$m, and the specific dust mass 
density of 3 g cm$^{-3}$. The mass of dust, $M_{\rm d}$ becomes 
\begin{equation}
 M_{\rm d} = 10^4\left( \frac{L}{10^8\ L_\odot}\right)
 \left(\frac{T_{\rm d}}{40\ \mathrm{K}}\right)^{-5}, 
\end{equation}
where $L$ is luminosity. Dust temperatures are set to be equal to
those derived from the above SED fitting. 

Table~\ref{dmasL} shows that the warm dust mass occupies less than 1\% of the total
dust mass in M101. The total gas-to-dust ratio is thus estimated to be
280, which is slightly larger than the accepted value of $100-200$ for
our Galaxy \citep{knapp}.  

\subsection{Spatial distributions of cold and warm dust}
In order to derive the distributions of the cold and warm dust 
components in the galaxy, the spatial resolutions of the WIDE-S and N60 
images are reduced to match those of the WIDE-L and N160 images by  
convolving the former images with a Gaussian kernel with the width of 20
arcsec, which has been performed before smoothing the images in figure~\ref{4image}.
The images are then resized with the common spatial scale among the four
bands: 25$\times$25 arcsec$^2$. As described in \S2,
the flux densities at each image bin are derived from the aperture
radius of 14 arcsec with the aperture-correction factors of 0.30 
for the four bands. 
An individual SED constructed from the four-band fluxes at each image bin 
is then fitted with a two-temperature model, in which the
temperatures are fixed at the values obtained for the SED of the whole
galaxy; we could not well constrain the dust temperatures from fitting
the four far-IR bands data if we set the temperatures to be free.  
Figure~\ref{cw_dust} shows the distributions of the dust emission thus
spectrally deconvolved into the two component. The map of the warm dust
is almost identical to the N60 image, which can be understood by
considering that the contribution of the cold dust component to the N60
band intensity is negligible as seen in the SED fitting of
figure~\ref{total_sed}. The distribution of the cold dust however shows
some differences from those of the other photometric band images as seen
in the giant HII regions, where the contributions of the two components
may be intermixed to some extent. 
 
As shown in figure 4, the cold dust component seems to be smoothly
distributed over the entire extent of the galaxy, while the warm dust
component indicates some correlation with the spiral arms. Figure 5
shows the R-band gray-scale image superposed on the contour map
of the cold dust component. The cold dust may be heated by a diffuse 
heating source, such as an old stellar population or non-ionising UV
photons escaping from H\emissiontype{II} regions. In outer regions,
the warm dust component is correlated with the spotty structures that
belong to the spiral arms. The morphology of their warm and cold dust
components seem to be consistent with those of other nearby galaxies
found by ISO observations (\cite{sauvage}) and confirmed by Spitzer 
observations (\cite{dale, hinz, gordon, taba}). The spots embedded in the
outer arms are spatially related to the four giant H\emissiontype{II}
regions as shown in the far-ultraviolet (UV) map with GALEX
(figure~\ref{comp2}). However, in the giant H\emissiontype{II}
regions NGC 5455, 5461, and 5462, the far-UV peaks are somewhat located  
at the peripheries of the spots in the warm dust emission. NGC 5461 is
located well outside the nuclear region but still carries a significant
fraction (1 \%) of the total far-IR luminosity of M101. This situation
is different from that in our Galaxy, where the far-IR emission from
dust associated with H\emissiontype{II} regions contributes only a minor
fraction; for example, the far-IR luminosity of the most active
star-forming region W49A is 0.2 \% of the total far-IR luminosity of our
Galaxy \citep{siever, bloemen}.    

The bar near the center was first discovered by CO $(J = 1-0)$ observations
\citep{kenney}, and the R-band image reveals an oval distortion in the
stellar distribution, which is offset in the position angle from the bar
by $\sim$25 arcsec. The AKARI/FIS shows that the warm dust component
traces the bar region, spanning 2 arcmin in length in the central region
of M101. This bar is not seen in the distribution of the cold dust, but can be traced also
in the far-UV emission \citep{gil}. Numerical simulations of interstellar
gas with barred potentials can produce this position angle offset due
to their dissipative nature \citep{hunt}; gas is concentrated on the
leading edge of the barred potential by a shock wave. Hence, we
conclude that the warm dust component traces the concentration of the
interstellar medium along the bar potential near the center, and that
there is active star formation not only in the center but also throughout the
bar. Near the end of the bar region labeled in figure~\ref{sfe_reg}, 
however, figure~\ref{comp2} shows significant deviations of both warm dust 
and far-UV distributions from the CO distribution; the warm dust emission 
is still bright beyond the leading edge of the CO emission, tracing the 
complex far-UV emission in the developed spiral arm. 
The difference in the spatial distribution between the warm dust emission and the CO emission may be
interpreted as evidence of star formation induced by a spiral density
wave \citep{rand, loinard}.

On the basis of the spectrally-deconvolved dust emission maps in figure 4, 
we derive the far-IR luminosity ratio between warm and cold dust
($L_\mathrm{w}/L_\mathrm{c}$) in various regions, which consist of seven
local fields in M101: the four giant H\emissiontype{II} regions, the
center, the bar, and the inter-arm region as shown in
figure~\ref{sfe_reg}. The results are shown in table~\ref{sfe}.  
It is found from the table that the ratios of the four giant H\emissiontype{II}
regions are significantly higher than those of the center and the bar,
and $20-40$ times as high as that of the inter-arm region. The cold dust
component is associated with gas content, while the warm dust component
is associated with the SFR of massive stars. Thus, the ratio
$L_\mathrm{w}/L_\mathrm{c}$ is physically related to star-formation
activity. Systematic errors in the ratio are likely to be dominated by the assumption of
constancy in the warm dust temperature; the dust temperature can be
higher than the fixed one particularly in the giant HII regions. However,
higher temperature results in larger $L_\mathrm{w}$, thus making the
difference in $L_\mathrm{w}/L_\mathrm{c}$ between the giant HII regions
and the other regions even larger. Therefore, the result in
table~\ref{sfe} may indicate that star-formation activity of the four
giant HII regions are highest in the galaxy. As a result of a past
encounter of M101 with its companion galaxy, NGC 5477, intergalactic HI
gas may fall in the outer disk near at least NGC 5461 and 5462
\citep{hulst}. Therefore, such external effects may promote the
formation of the giant H\emissiontype{II} regions. As for the bar, the
ratio of the entire bar is similar to that of the center. 

\subsection{Color temperature of dust in various regions}
Since it is not easy in practice to accurately estimate $T_\mathrm{c}$ and
$T_\mathrm{w}$ for each image bin, temperatures of the cold (18 K) and warm (55 K)
dust components are kept to be constant over the galaxy for simplicity in the 
above discussion. 
Nevertheless, the far-IR colors constructed from any combination of the FIS 
four-band fluxes can be robust indicators of the temperatures of the warm and 
cold dust for each region in M101. Figure~\ref{color_color} shows the 
color-color diagram log(F140/F160) versus log(F65/F90) of the above seven 
local fields and the whole galaxy in figure~\ref{total_sed} by using an 
average over each region. F65, F90, F140, and F160 show the flux
densities of the N60, WIDE-S, WIDE-L, 
and N160 bands, respectively. In figure~\ref{color_color}, the F65/F90
ratio is significantly higher in the four giant H\emissiontype{II}
regions, the center, and the bar as compared to that of the 
galaxy as a whole. The higher F65/F90 colors can be interpreted as higher 
warm dust temperatures, i.e. more active star formation. Therefore,
star-formation activities in the six regions are more intense than those
in the rest of the galaxy.  
In particular, the four giant H\emissiontype{II} regions show the most 
active star formation in M101. Hence, the results show overall consistency 
with the ratio $L_\mathrm{w}/L_\mathrm{c}$ listed in table~\ref{sfe}.         

On the other hand, the F140/F160 colors do not show significant
variations among various regions except NGC 5455. Since the F140/F160 color
can be related to cold dust temperature, variations in cold dust temperature are 
significantly smaller than that in warm dust temperature over the
galaxy. This fact is consistent with the idea derived by \citet{shibai},
which also justifies the above assumption of constant temperature for
cold dust. One exception is NGC 5455, which shows significantly higher F140/F160
color than the other giant H\emissiontype{II} regions. As shown in
figure~\ref{metallicity}, the oxygen abundance [12+Log(O/H)] in the four
giant H\emissiontype{II} regions is lower $(35-60 \%)$ than that of the solar
system \citep{kennicut3}; metallicity in NGC 5455 is the lowest among
the four giant H\emissiontype{II} regions. \citet{madden} present that
low metallicity regions indicate intense ISRF. Therefore the higher
F140/F160 color of NGC 5455 may be explained by more intense ISRF in NGC 5455.   

Although the IRAS observation have unexpectedly shown that the
far-IR colors are insensitive to the change of the ISRF intensity
over the galaxy \citep{nicholas}, the AKARI/FIS has demonstrated 
that there are significant variations in the far-IR colors among the seven 
regions with the four photometric bands and high spatial resolution. 
Although M101 is classified as an inactive galaxy, unlike our Galaxy, 
the extraordinary active star-forming regions are located in the outer disk
($10-16$ kpc from the center; \cite{israel}). M101 is thus considered
to be a peculiar inactive galaxy.    

\section{Conclusions}
The spatial structure of M101 is well resolved in the four bands with the
AKARI/FIS. The resultant SED of the whole galaxy shows the presence of a cold
dust component (18$^{+4}_{-9}$ K) in addition to a warm dust component
(55$^{+9}_{-25}$ K). Considering its small
$L_\mathrm{FIR}/M_\mathrm{gas}$ value of 0.9 $\LO/\MO$, M101
is classified as an inactive galaxy.  
We have deconvolved the cold and warm dust emission components spatially, 
by making the best use of the multi-band photometric capability of the FIS.
The difference in the distribution between the cold and warm dust is 
more clearly revealed than in the separate images in each
of the four bands. The distribution of the cold dust is mostly concentrated
near the center, and exhibits smoothly distributed over the entire extent of the galaxy,
whereas the distribution of the warm dust indicates some correlation
with the spiral arms, and has spotty structures such as four distinctive
bright spots in the outer disk in addition to a bar-like feature near
the center tracing the CO intensity map. The former bright spots are spatially
related to the four giant H\emissiontype{II} regions. The latter feature
implies that there is active star formation throughout the bar filled
with molecular clouds.

On the basis of the distributions of the warm and cold dust
components, we have derived $L_\mathrm{w}/L_\mathrm{c}$ as a robust
measure of star-formation activity in the various regions in
M101. Star-formation activity of the four giant H\emissiontype{II} 
regions is significantly higher than that of the rest of the galaxy. External effects
such as the infall of intergalactic gas may promote the formation of the
four giant H\emissiontype{II} regions. The color-color diagram
constructed from any combination of the FIS four-band fluxes has
revealed the temperature variations of the warm and cold dust in the
various regions. The four giant H\emissiontype{II} regions show
significantly higher warm dust temperatures in M101, which supports the
active star formation. The warm dust temperature shows large 
variations over the galaxy, whereas the cold dust temperature shows
a comparatively narrow range of variations, which suggests relatively 
constant ISRF intensities over the galaxy. Hence, unlike our Galaxy, 
M101 is considered to be a peculiar inactive galaxy
with extraordinary active star-forming regions.

\bigskip
We would like to thank all the members of the AKARI project for their
continuous help and support. Based on observations with AKARI, a JAXA
project with the participation of ESA. We are grateful to the AKARI data
reduction team for their extensive work in developing data analysis pipelines.
The research presented in this paper used data from the National
Geographic Society$-$Palomar Observatory Sky Atlas (POSS-I) that was
made by the California Institute of Technology with grants from the
National Geographic Society.

\clearpage

\begin{figure}
  \begin{center}
    \FigureFile(80mm,80mm){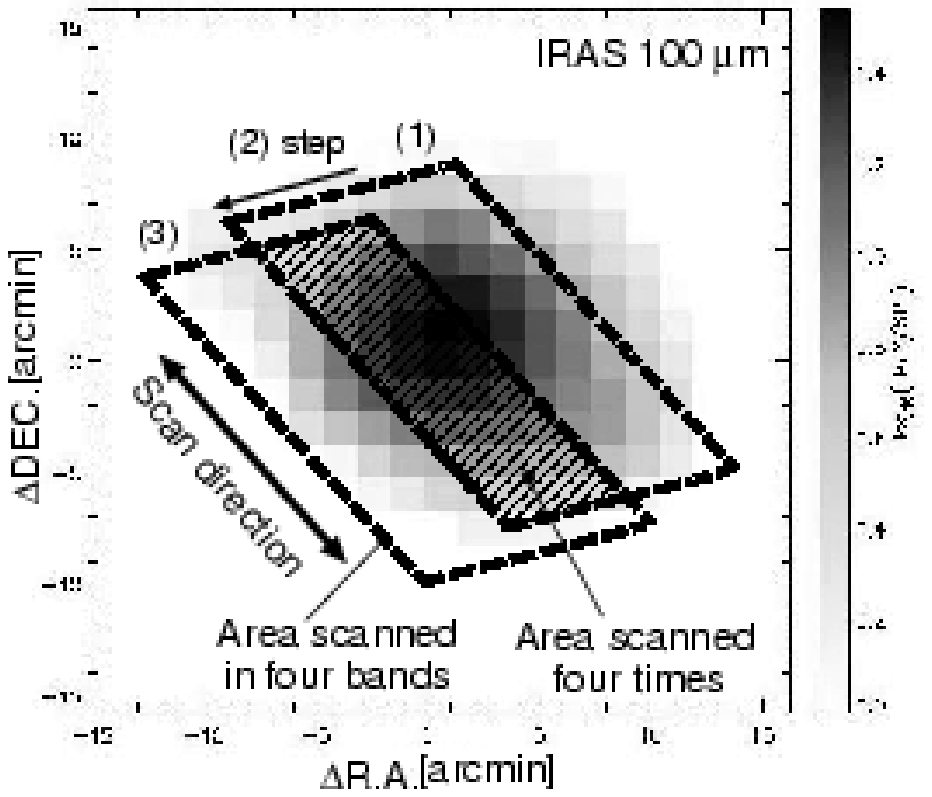}
  \end{center}
  \caption{Area covered in the four far-IR bands with the FIS slow scan
  observations. The gray image is taken from the IRAS 100 $\mu$m
  map. Wider area is obtained at the direction of the scan every
  photometric band.} 

\end{figure}

\begin{figure}[t]
  \begin{center}
    \FigureFile(160mm,80mm){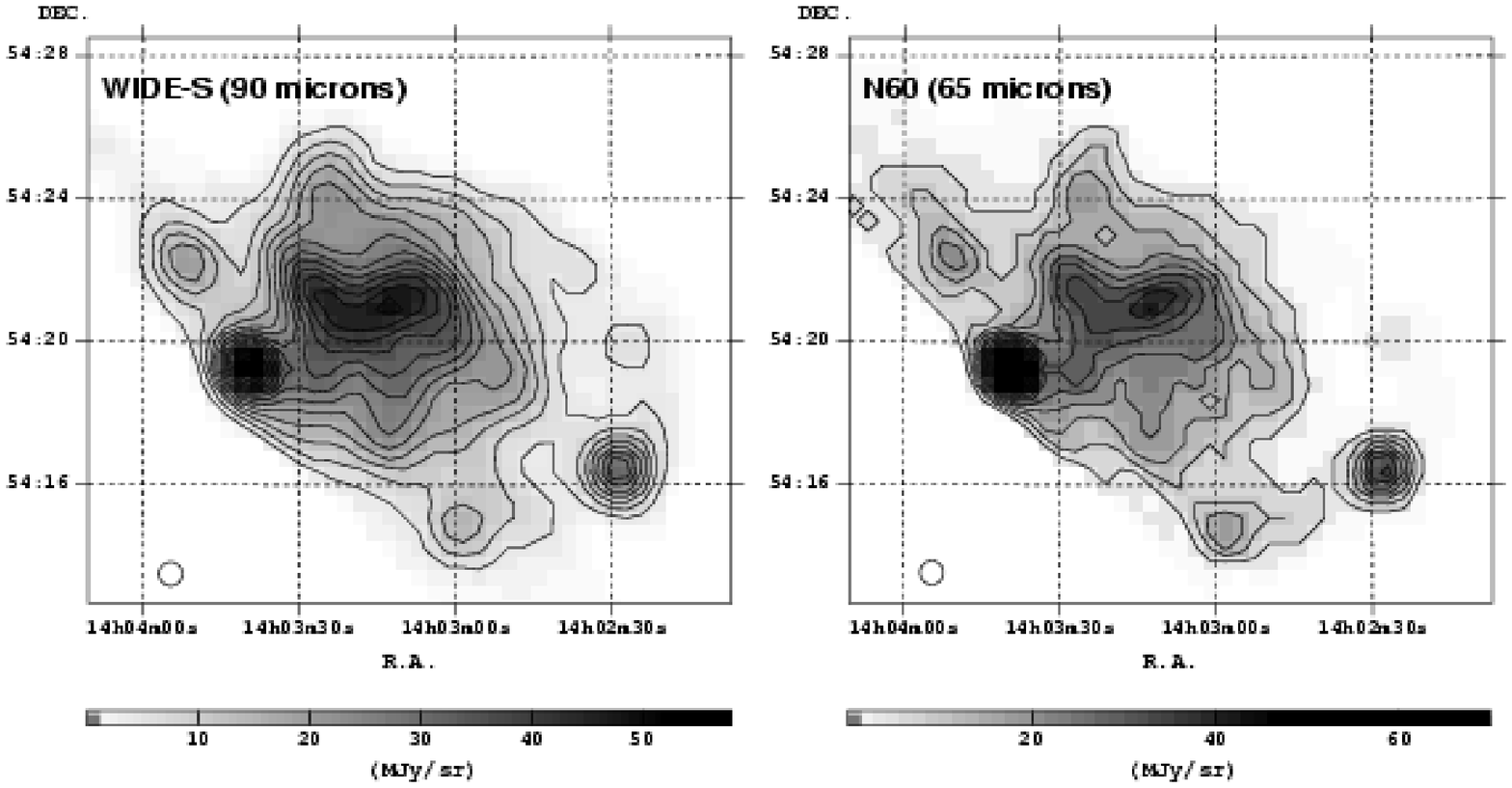}
    \FigureFile(160mm,80mm){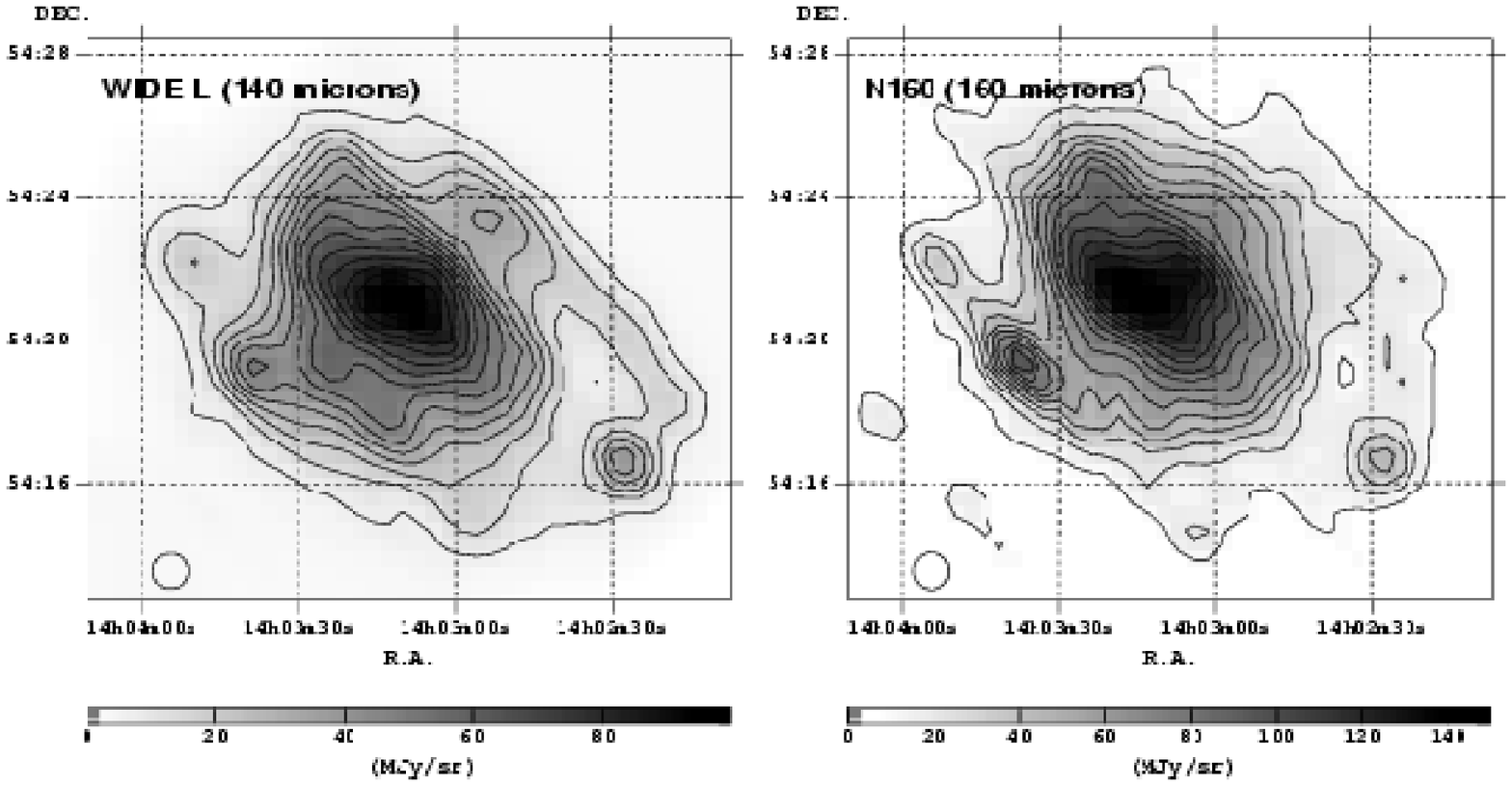}
  \end{center}
  \caption{Four-band images of M101 in the WIDE-S (top-left), N60 (top-right), WIDE-L (bottom-left), and
 N160 (bottom-right) bands. The center wavelengths of the four bands are 65
 $\mu$m for N60, 90 $\mu$m for WIDE-S, 140 $\mu$m for WIDE-L, and 160
 $\mu$m for N160. Contours are linearly spaced from 7 \% to 98 \% of the
 peak brightness at a step of 7 \%. Peak brightness is 60
 MJy/sr (WIDE-S), 70 MJy/sr (N60), 100 MJy/sr (WIDE-L), and 150 MJy/sr
 (N160). Typical 1$\sigma$ noise levels are 0.1 MJy/sr (WIDE-S), 0.4 MJy/sr (N60), 0.1 MJy/sr
 (WIDE-L), and 0.3 MJy/sr (N160). In each image, the PSF size in FWHM is
 shown in the lower left corner.} 
\label{4image} 
\vspace{3mm}
\end{figure}

 \begin{figure}
\begin{center}
    \FigureFile(90mm,95mm){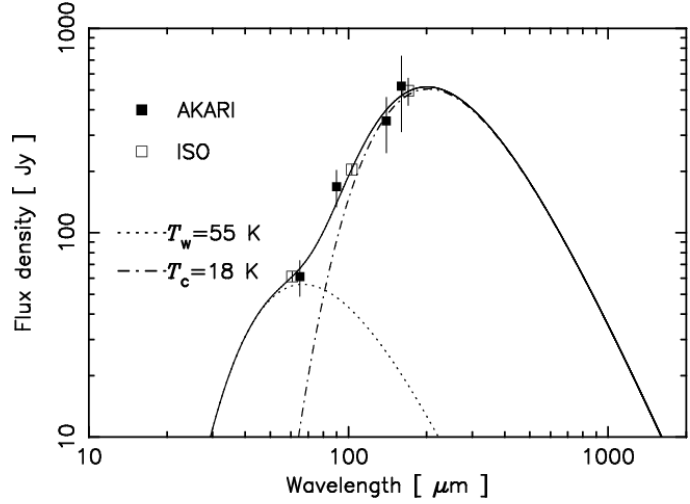}
\end{center}
  \caption{Spectral energy distribution of M101, together with the
 best-fit double-temperature modified blackbody model. Filled boxes
 represent the integrated flux densities in the FIS four bands, while
 open boxes correspond to those in the ISOPHOT far-IR
 bands \citep{tuff}. }\label{total_sed}  
\end{figure}

\begin{figure}[t]
  \begin{center}
    \FigureFile(160mm,80mm){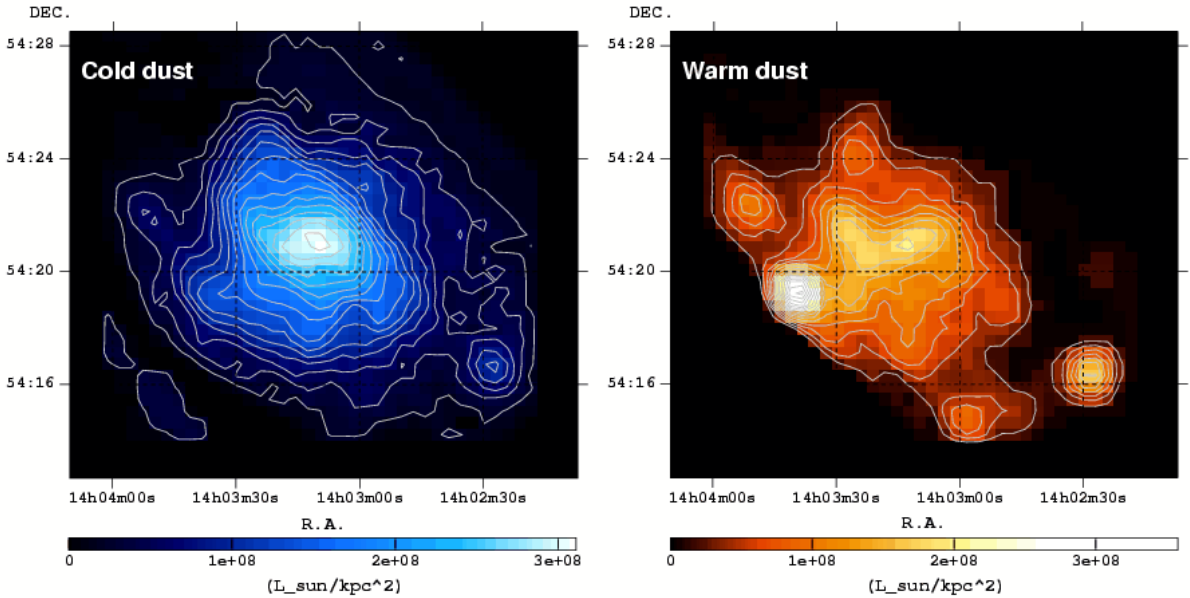}
    \end{center}
  \caption{Spatial distributions of the cold dust (left), and warm dust (right)
 components of M101. Contours are linearly spaced from 7
  \% to 98 \% of the peak at a step of 7 \%. Peak luminosity
  is 3.6$\times10^8$ \LO/kpc$^2$ for the warm dust component, and
  3.1$\times10^8$ \LO/kpc$^2$ for the cold dust component.} 
\label{cw_dust} 
\vspace{3mm}
\end{figure}

\begin{figure}
  \begin{center}
    \FigureFile(80mm,80mm){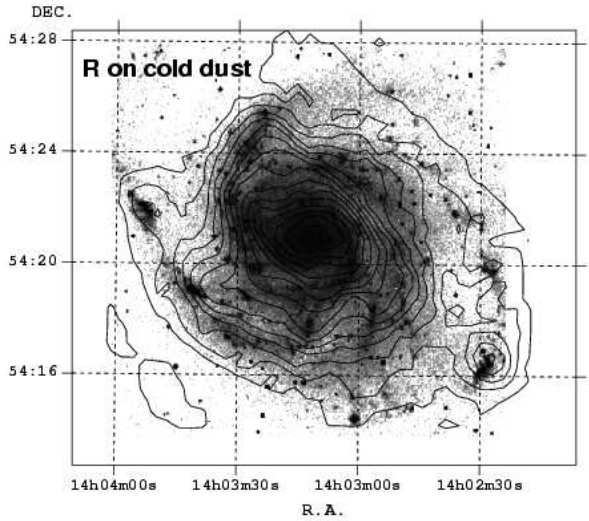}
  \end{center}
 \caption{R-band gray-scale image (DSS) of
 M101 superposed on the contour map of the cold dust emission. The
 contour levels are the same as in figure~\ref{cw_dust}.} 
 \label{comp}
\vspace{3mm}
\end{figure}

\begin{figure*}
  \begin{center}
    \FigureFile(80mm,80mm){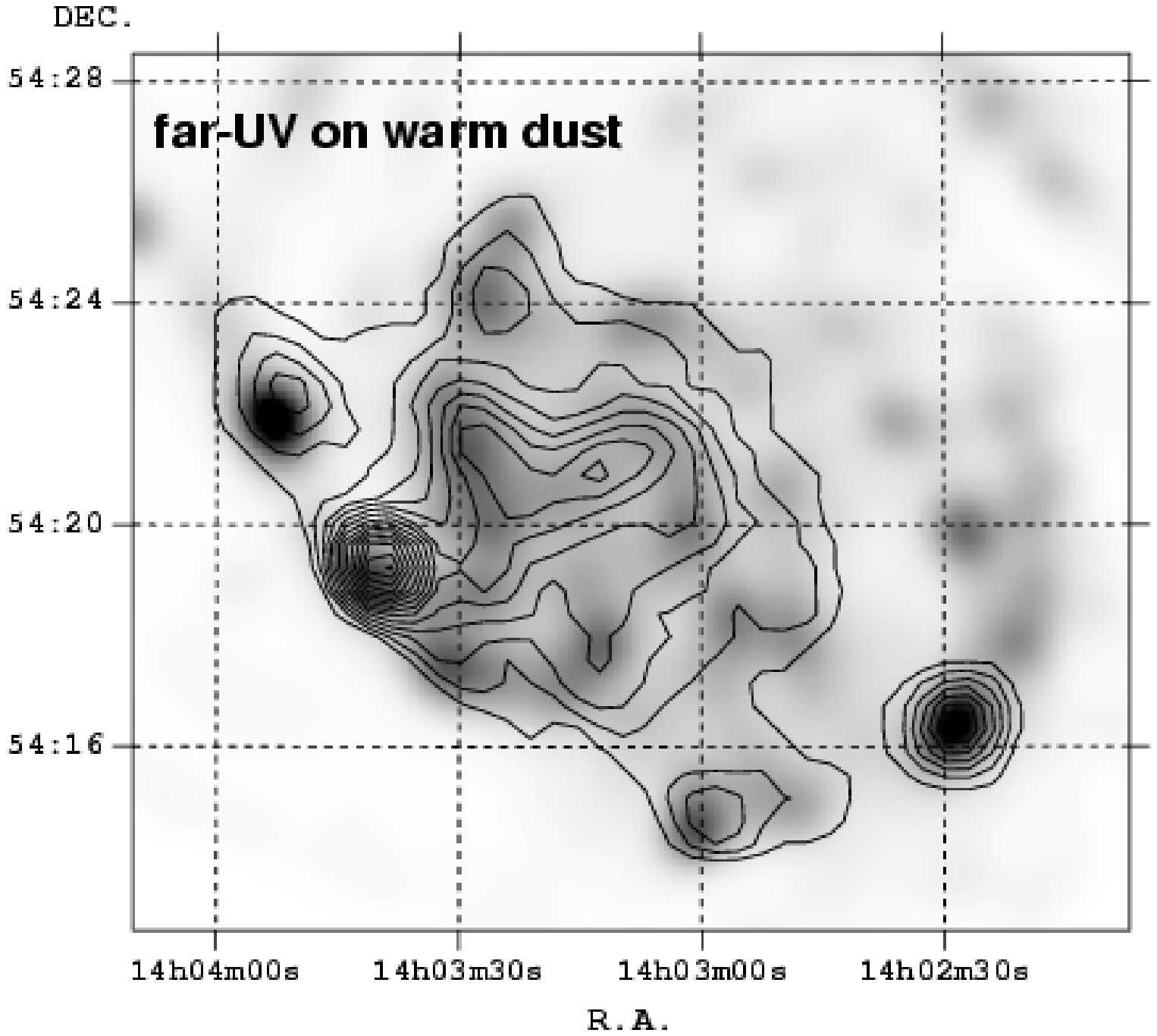}
    \FigureFile(80mm,80mm){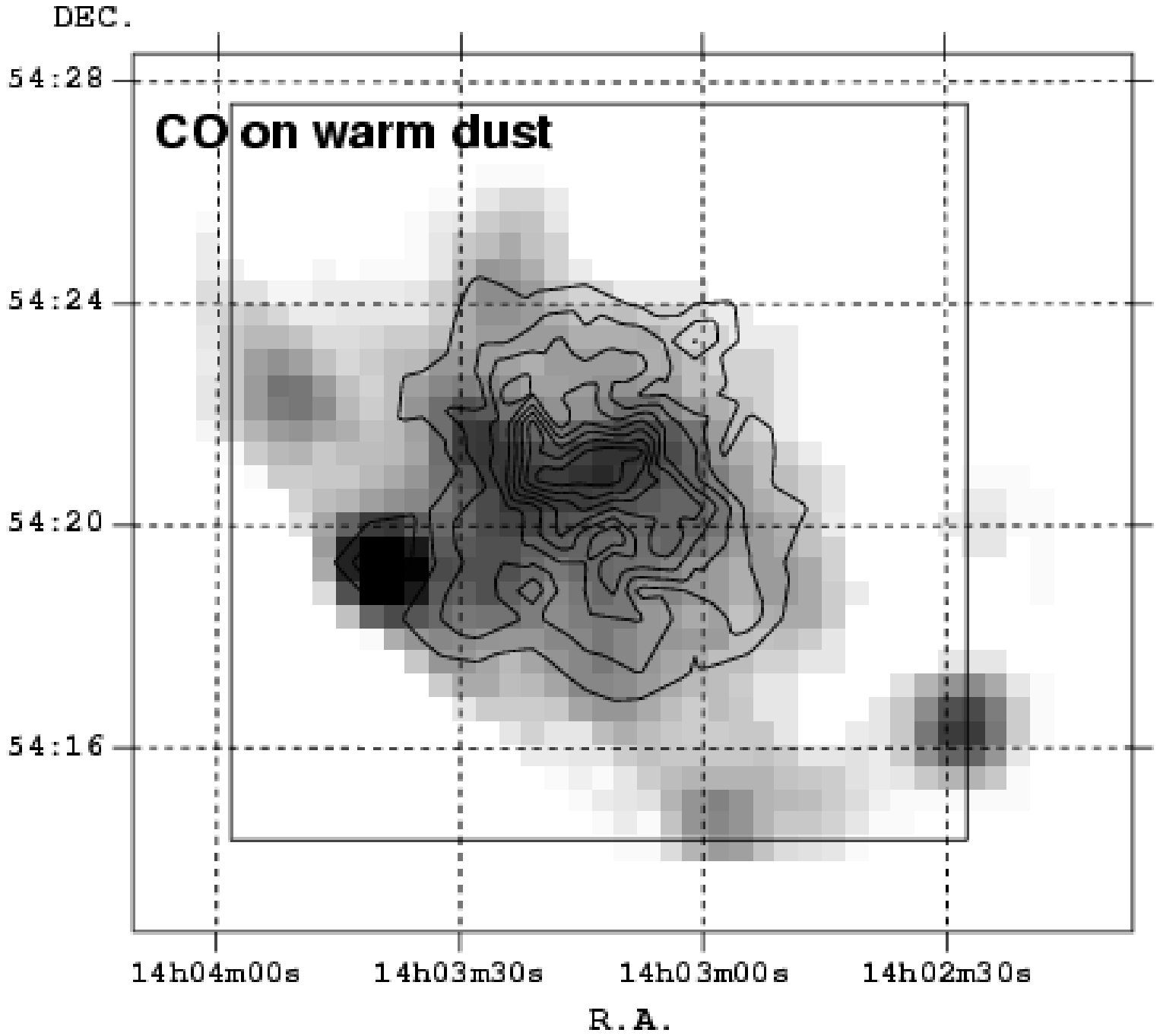}
  \end{center}
 \caption{(left) Far-ultraviolet ($\lambda=$1516 \AA) gray-scale image
 of M101 \citep{gil} overlaid with the contour map of the warm dust
 emission. The far-UV image is convolved with the beam size same as
 in figure 4. The contour levels are the same as in
 figure~\ref{cw_dust}. (right) CO contour map of the M101 \citep{kenney}
 with the gray-scale map of the warm dust emission. The open box shows the area of the
 CO observation.}   
 \label{comp2}
\vspace{3mm}
\end{figure*}

\begin{figure}
  \begin{center}
    \FigureFile(80mm,80mm){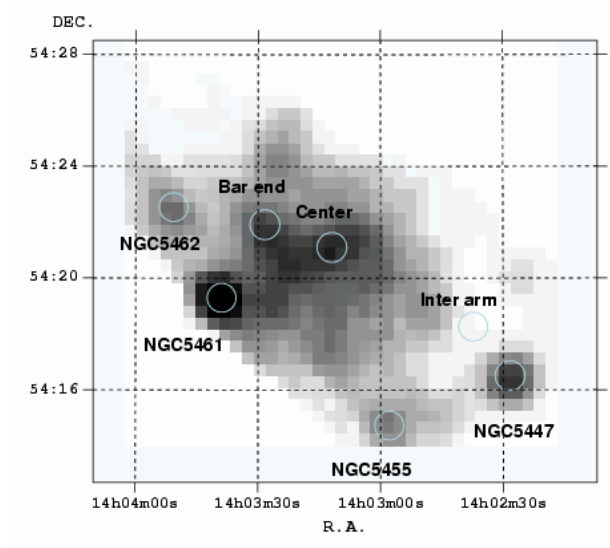}
  \end{center}
 \caption{Positions of seven local fields in M101 indicated as the
 circles in the warm dust emission map (the same as the right panel of
 figure~\ref{cw_dust}); for each region, the luminosity of the warm dust
 emission, the dust mass of the cold dust, and thus the ratio
 $L_\mathrm{w}/L_\mathrm{c}$ is obtained (see text for details)} 
 \label{sfe_reg}
\vspace{3mm}
\end{figure}

\begin{figure}
 \hspace{2mm}
\begin{center}
    \FigureFile(94mm,94mm){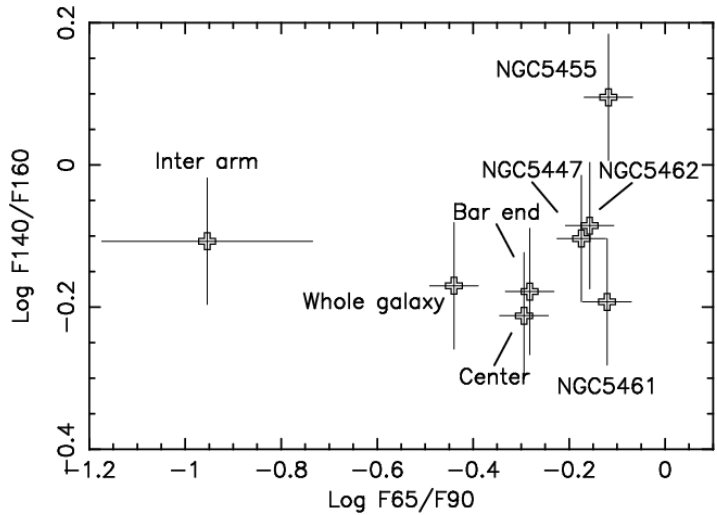}
\end{center}
  \caption{Color-color diagram of various regions in M101. F65, F90, F140,
    and F160 correspond to the flux densities of the N60, WIDE-S, WIDE-L and N160
    bands, respectively.}\label{color_color}
 \end{figure}

  \begin{figure}
\begin{center}
    \FigureFile(94mm,94mm){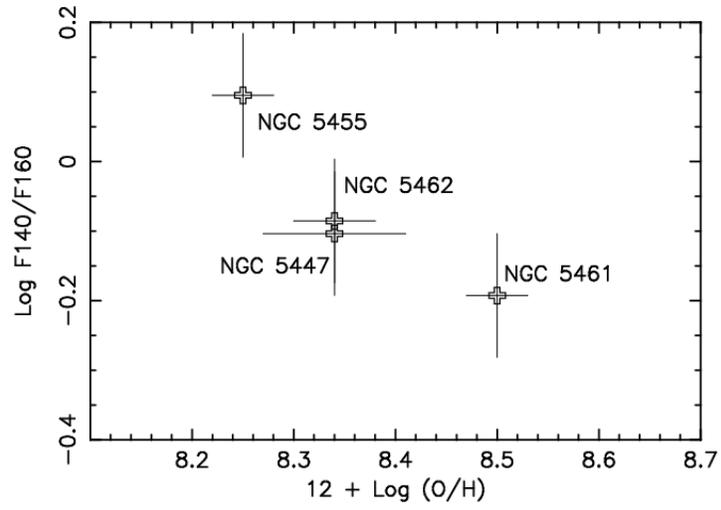}
\end{center}
 \caption{Oxygen abundance \citep{kennicut3} versus the F140/F160 color
 in the four giant H\emissiontype{II} regions. The solar oxygen
 abundance [12 + log(O/H)] is  8.70 \citep{allende}. }
 \label{metallicity}
\vspace{5mm}
\end{figure}

\clearpage
\begin{table*}
\vspace{10mm}
  \caption{Flux densities of M101}\label{flux}
 \vspace{-2mm}
 \begin{center}
  \begin{tabular}{cccc}   
    \hline \hline 
    N60 (65 $\mu$m) & WIDE-S (90 $\mu$m)& WIDE-L (140 $\mu$m)& N160 (160 $\mu$m)\\ \hline
    61$\pm12$ Jy & 168$\pm$34Jy & 353$\pm$106 Jy & 522$\pm$209 Jy \\  \hline 
  \end{tabular} 
 \end{center}
\end{table*}

\begin{table*}
\vspace{10mm}
  \caption{Properties of the far-infrared dust emission in M101}\label{dmasL}
 \vspace{-2mm}
 \begin{center}
  \begin{tabular}{ccc}   
  \hline \hline 
   &  Far-IR luminosity ($L_\odot$) & Dust mass ($M_\odot$)    \\ \hline 
   Cold dust & (1.6$^{+0.5}_{-0.4}$)$\times 10^{10}$ &
   (9$^{+90}_{-5}$)$\times 10^7$ \\  
   Warm dust & (5.3$^{+4.7}_{-0.6}$)$\times 10^{9}$ &
   (1.1$^{+40}_{-0.5}$)$\times 10^5$ \\
   \hline 
  \end{tabular} 
 \end{center}
\end{table*}

\begin{table*}
\vspace{3mm}
  \caption{Far-infrared luminosity ratio between cold and warm dust components
  ($L_\mathrm{w}/L_\mathrm{c}$) in various regions in M101}\label{sfe}
 \vspace{-2mm}
 \begin{center}
   \begin{tabular}{cccccccc}   
     \hline \hline 
            &Center &Bar end& NGC 5447 &
            NGC 5455 & NGC 5461 &
            NGC 5462 & Inter arm \\ \hline

$L_\mathrm{w}/L_\mathrm{c}$ 
            &0.6${\pm0.1}$ &0.8${\pm0.2}$ &
            1.6${\pm0.4}$ & 3.0${\pm0.7}$&
            2.6${\pm0.6}$& 1.9${\pm0.4}$& 0.07${\pm0.02}$\\  \hline
 \end{tabular} 
 \end{center}
\end{table*}

\end{document}